\documentclass[12pt]{article}
 
\newcommand{\vs}{\vspace{15pt}} \newcommand{\n}{\noindent} 
\newcommand{\rf}[1]{(\ref{#1})}
\newcommand{\ba}{\begin{array}} \newcommand{\ea}{\end{array}}
\newcommand{\be}{\begin{equation}} 
\newcommand{\btb}{\begin{tabular}}\newcommand{\etb}{\end{tabular}}
\newcommand{\ee}[1]{\label{#1}\end{equation}}
\newcommand{\bi}{\bibitem} 
\newtheorem{thm}{Theorem}[section] \newtheorem{pro}[thm]{Proposition}
\newtheorem{cor}[thm]{Corollary}
\newtheorem{df}[thm]{Definition} 
\newcommand{\dss}{\displaystyle}
\newcommand{\bfl}{\begin{flushleft}}\newcommand{\efl}{\end{flushleft}}

\newcommand{\al}{\alpha} \newcommand{\bt}{\beta}
\newcommand{\g}{\gamma}\newcommand{\G}{\Gamma}  

\newcommand{\ep}{\epsilon}
  
\newcommand{\la}{\lambda}\newcommand{\La}{\Lambda}  
\newcommand{\si}{\sigma}\newcommand{\Si}{\Sigma}

  \newcommand{\R}{\mathbb R}  
 
\newcommand{\EC}{{\cal E}}\newcommand{\FC}{{\cal F}}
\newcommand{\GC}{{\cal G}}

\newcommand{\VC}{{\cal V}}\newcommand{\XC}{{\cal X}}

\newcommand{\LCU}{{\cal L}_\uparrow} 
\newcommand{\PCU}{{\cal P}_\uparrow} 

\newcommand{\Ay}{{\bf A}} \newcommand{\By}{{\bf B}}
\newcommand{\Cy}{{\bf C}} 
 \newcommand{\Ky}{{\bf K}}
\newcommand{\Uy}{{\bf U}} 
\newcommand{\Vy}{{\bf V}} \newcommand{\ay}{{\bf a}}\newcommand{\by}{{\bf b}} 
\newcommand{\ey}{{\bf e}}

\newcommand{\ky}{{\bf k}}

\newcommand{\ry}{{\bf r}}\newcommand{\sy}{{\bf s}} 
\newcommand{\uy}{{\bf u}}\newcommand{\vy}{{\bf v}} 
\newcommand{\0}{{\mathbf 0}}  
\newcommand{\Wy}{{\bf W}}

\newcommand{\dof}{\dot{f}}

\newcommand{\hal}{\hat{\alpha}}

\newcommand{\Dia}{\Diamond}
\newcommand{\na}{\nabla} \newcommand{\p}{\partial}  
  
\newcommand{\ra}{\rightarrow} 
\newcommand{\ol}{\overline}  
\newcommand{\sq}{\subseteq}

 \newcommand{\vers}{\mbox{vers}}

\newcommand{\diag}{\mbox{diag}}\newcommand{\Int}{\mbox{Int}}
\newcommand{\Bi}{\mbox{Bi}}

\usepackage{amssymb}
\usepackage[T1]{fontenc}

\title{On the Fundamental Theorem of the Theory of Relativity}
\author{Marco \textsc{Mamone-Capria}\\  \small Dipartimento di Matematica --  via Vanvitelli, 1 --  06123 Perugia - Italy \\ \small {\sl E-mail}: 
 \texttt{mamone@dmi.unipg.it} }

\begin{document}
\maketitle
\begin{abstract}
A new formulation of what may be called the ``fundamental theorem of the theory of relativity'' is presented and proved in (3+1)-space-time, based on the full classification of special transformations and the corresponding velocity addition laws. A system of axioms is introduced and discussed leading to the result, and a study is made of several variants of that system. In particular the status of the group axiom is investigated with respect to the condition of the two-way isotropy of light. Several issues which are ignored or misunderstood in the literature are emphasized.  

\vs

\end{abstract}

\tableofcontents

\section{Introduction}

The discovery of the equations of the special Lorentz transformation (SLT) prompted several scholars, since the early days of the special theory of relativity, to look for general assumptions, possibly less questionable than the postulate of constancy of the velocity of light, from which those equations could be derived. In 1936 the Italian mathematician Francesco Severi echoed a widespread concern when he qualified the ``intervention of the light-velocity'' in special relativity as ``the scandal of the theory'' (\cite{s36}, p. 260). The search for alternative assumptions was on the whole successful, and in fact produced several derivations of the SLT --  and, at the same time, of the corresponding classical, Galilean transformations --  from general axioms on space and time taken together with some `relativity' condition. This search represents an important, if somewhat neglected, undercurrent in the development and reception of special relativity, which has attracted very many physicists and mathematicians since the pioneering contributions by von Ignatowsky (\cite{i10a, i10b}) till the present day.

What I call the `Fundamental Theorem of the Theory of Relativity' is an ideal theorem which puts together the essential insights gained through this volume of work. In the present paper I advance one proposal for such a theorem -- the kind of statement that, accompanied by physical and historical supporting arguments, should  find its way into textbooks. The basic idea is not new in itself: {\sl to create an appropriate formal and conceptual environment which allows for the notion of special transformation to develop into a structure group}.      

Too many articles on this subject have been published in more than a century for a single scholar to be sure that nothing relevant escaped his or her attention. I made a honest attempt not to miss apparently important texts, my list of references being a representative sample of the (often repetitious) literature I checked. I have come to the conclusion that many of the arguments found in the  literature are flawed, unsatisfactory, or incomplete. One reason for my dissatisfaction is that those arguments usually and tacitly assume that the passage from the derivation in (1+1) space-time to the derivation in (3+1) space-time is straightforward, which is not; also, in most treatments the notion of a physical process or of signalling is not sufficiently clarified; and, to give a further example, the status of the group assumption is often left hazy. 

Still, in a classical field such as that of the foundations of relativity it would be hazardous, if not arrogant, to claim an high degree of originality for one's results. Several authors have found themselves informed by colleagues in due time, and sometimes shortly after publication, that what they had published was close, and sometimes very close, to results already in the published record.\footnote{See for instance the ``Postscript'' to \cite{mer84} added in \cite{mer} or the ``Note added in proof'' of \cite{ll76}. In some cases, like \cite{bg69} vs. \cite{c58}, the `anticipated' paper has come to be much more famous and cited than the `anticipating' one.} I have done my best to avoid this risk, by screening the literature for similarities with my main results. 

In \S 2 the basic definitions and axioms are stated concerning the class of privileged, or admissible, coordinate systems (space-time structure); they make, first of all, the space-time into a differential manifold diffeomorphic to standard $\R^4$. The concept of a worldline is introduced, together with axioms guaranteeing that all admissible coordinate systems share the same notion of causality, time orientation and spatial orientation. One axiom which is rarely, if ever, explicitly stated in the literature is Axiom 6, which limits the richness of the space-time structure by requiring that mutually at rest cs's have the same spatial geometry and, up to an additive constant, the same time coordinate. Of course this rules out all unit changes but, more importantly, prevents the introduction of nonstandard synchronies. These are dealt with separately in terms of the same theoretical framework, and their relationship with the group condition (Axiom 7) is examined  in \S 6. 

Inertial equivalence is carefully defined in \S 3, in a way which takes into account, as is unfrequently done, the fact that in a theory which is designed to include as particular cases both classical mechanics and special relativity not all velocities may be allowed a priori for uniform motions. The link between inertial equivalence and affine equivalence is stressed, with a criticism of the usual approach to it in terms of `homogeneity of space-time'. Axiom 10 links the set of physical velocities with the admissible coordinate systems. Section 4 is devoted to the concept of special transformation, which is at the core of the vast majority of treatments of the `fundamental theorem', and contains a theorem which reaches back to Frank and Rothe's 1911 classic paper (\cite{fr11}), but which seems never to have been fully stated and proved. It is also shown that not all space-time structures satisfying axioms 1-9 satisfy Axioms 10 and 11, in particular they may not admit any one-dimensional subgroups of special transformations. In the next section the final condition, spatial isotropy (Axiom 12), and my proposal for the `fundamental theorem' are presented, the latter assuming all axioms 1-12. The full family of trasformations satisfying two-way light velocity isotropy (Axiom $6^\ast$) is derived in \S 6, and the issue of the compatibilty of  Axioms $6^\ast$ and 7 is discussed,  which motivates the introduction of what I call the Reichenbach groups, with some comments on their relationship to the issue of the conventionality of simultaneity.

\section{Basic assumptions (Axioms 1-7)}

The general framework adopted in this paper is essentially the traditional one, reconstructing space-time structures by means of classes of admissible (or privileged) coordinate systems. A space-time theory is specified when the rules for the selection of such coordinate systems are formulated. This contrasts with the approach favored in purely mathematical treatments, where geometric structures are superimposed from the start to some set, and privileged coordinate systems are derived thence. In our approach, for instance, the requirement that `space-time has a real affine structure' is to be reformulated in terms of the existence of a class of affinely equivalent coordinate systems; and a reasonably physical, if idealized, criterion must be provided for determining whether a given coordinate system belongs or not to such a class. We denote by $\EC$ the {\sl space-time}; its elements are called {\sl events}. In the following only the more important items (definitions, theorems) are numbered; axioms are listed and numbered separately.

\begin{df} A {\bf (global) coordinate system} (cs) is a bijection $\phi: \EC \ra \R^4$. A {\bf space-time structure}  is a nonempty class $\Phi$ of global coordinate systems, which are variously termed as {\bf privileged}  or {\bf admissible}. A cs in the space-time structure $\Phi$ is denoted  by

\[ \phi : \EC\ra \R^4 \equiv \R^3 \times \R, \; p \mapsto (\ry (p), t(p)) , \]

\n
where $\ry\equiv \ry_\phi : \EC \ra \R^3$ is the {\bf position function} and $t\equiv t_\phi : \EC \ra \R$ is the {\bf time function}  of $\phi$. A property or a quantity is {\bf absolute} if it is the same with respect to all $\phi\in\Phi$; it is {\bf relative} otherwise. Two events $p,q$ are {\bf synchronous}  with respect to $\phi$ if $t (p) = t(q)$.\end{df} 

By {\sl theory of relativity} I mean the theoretical framework which is described by {\sl axioms 1-5} only. Several other axioms will be stated in due course producing different specializations of the theory. I will not insist on the considerable amount of idealization in each of the axioms, starting with the first, which is a cardinality axiom.

\n
\textbf{\textsc{Axiom 1} [space-time structure]} {\sl  Physical space-time has a space-time structure $\Phi$}.

\n
\textbf{\textsc{Axiom 2} [topology and differential structure]}   {\sl  For every $\phi, \phi'\in \Phi$ the transition function 

\[ \phi'\circ\phi^{-1} : \R^4 \ra \R^4, \; \mbox{or equivalently} \; \left\{\ba{rcl} \ry'&=& \ry'(\ry, t) \\ t'&=& t'(\ry, t) \ea\right. , \]

\n 
is a $C^2$ diffeomorphism. Space-time $\EC$ is endowed with the topological and differential structures determined by any atlas consisting of a single global chart $(\EC, \phi)$ with $\phi$ admissible and  $\R^4$ standard}.

That $\EC$ be topologically and differentially modeled on standard 4-dimensional Euclidean space is a nontrivial assumption that is maintained, in local form, also in general relativity. In particular, from the assumption that $\EC$ is homeomorphic to $\R^3 \times \R$ (with the product topology) it does not follow that $\EC$ is also diffeomorphic to $\R^3 \times \R$ (with the product differential structure).\footnote{I remind the reader of the highly remarkable fact that of all $\R^n$s, $\R^4$ is the only one to have non-diffeomorphic differential structures; indeed, there is a {\sl continuum} of such structures (for an introduction to these results in differential topology, see for instance chapter I of \cite {bl85}).} 

A crucial concept is that of a physical process in space-time, formally introduced as follows.

\n
\begin{df} A {\bf worldline} is any regular curve in $\EC$ which can be totally parametrized by $\phi$'s time coordinate, for every $\phi\in\Phi$.
\end{df} 

This is to be understood as follows: a worldline $\G$ is such that for every $\phi \in\Phi$ there is a differentiable map \( \ry : I \ra \R^3\) such that $\phi (\G)$ is the inverted graph of $\ry$, i.e. $\phi (\G) = \{ (\ry (s), s) \; : \; s\in I\}$, where $I$ is a (nondegenerate) interval of $\R$; we say that $\ry$ {\sl represents} $\G$ in the 3-space of $\phi$. Given such a function, the 3-vector 

\[ \vy := \frac{d\ry}{ds}, \]

\n
is the {\sl 3-velocity} (i. e. the ordinary physical velocity) of the worldline at a fixed event according to $\phi$. \footnote{For other purposes, which will not concern us here, one may admit nonregular curves as worldlines, e.g. with a discrete subset of points at which 3-velocity is not well-defined.}

\n
\begin{df}A {\bf point at rest} in any admissible cs $\phi$ is the image of a map of the form  

\[ \g: \R \ra \EC, \; s\mapsto \phi^{-1} (\ry_0, s). \]

\n 
where $\ry_0\in \R^3$.\end{df}

\n
\textbf{\textsc{Axiom 3}}  [{\bf causality}] {\sl Any point  at rest in an admissible cs is a worldline}. 

This is equivalent to requiring the following condition to hold:

\be \frac{\p t'}{\p t} \neq 0 \ee{cc}

\n
for every pair of admissible $\phi, \phi'$ and for all $(\ry, t)\in \R^4$; when inequality \rf{cc} holds, $\phi$ and $\phi'$ may be seen as `causally compatible'. 

\n
\textbf{\textsc{Axiom 4}}  [{\bf time orientation}] {\sl All admissible cs's define the same time order on all worldlines.} 

Given \rf{cc}, this is equivalent to requiring the following condition to hold:

\be \frac{\p t'}{\p t} > 0 \ee{cto}

It is easy to verify, by exchanging roles between $\phi$ and $\phi'$, that as a consequence the spatial Jacobian determinant must be nowhere zero:

\be \det (\frac{\p x'^\al}{\p x^\bt})_{\al, \bt =1,2,3}\neq 0. \ee{spor}

\n
\textbf{\textsc{ Axiom 5}} [{\bf spatial orientation}] {\sl All admissible cs's define the same spatial orientation (i. e., the determinant in \rf{spor} is positive, for every pair of admissible cs's)}.

Because of Axiom 3, spatial orientation is equivalent to space-time orientation, thus for all admissible $\phi, \phi'$, the transition function $\phi'\circ\phi^{-1}$ is an orientation-preserving diffeomorphism of $\R^4$. 

Here ends the list of the few basic axioms defining our most general version of theory of relativity (which, of course, fails to encompass ``general relativity'', with or without the field equation).

The coordinates of admissible cs's admit of a natural interpretation. If $p$ and $q$ are $\phi$-{\sl synchronous}, or $\phi$-{\sl simultaneous}, events (i.e. $t(p) = t(q)$), their {\sl $\phi$-distance} is

\[ d_\phi (p,q) := |\ry (p)-\ry (q)|. \]

\n
On the other hand, for a point at rest in $\phi$ the difference of the $\phi$-time coordinates of two events in the worldline is interpreted as the (positive or negative) time lapse between the events as measured by a clock described by that worldline. 

Given admissible cs's $\phi$ and $\phi'$, we say that $\phi'$ is {\sl at rest} with respect to $\phi$ if all points at rest in $\phi'$ are also at rest in $\phi$. It is easy to see that if $\phi'$ is at rest with respect to $\phi$, then also $\phi$ is at rest with respect to $\phi'$ (so ``reciprocity'' -- cf. \S 5 -- holds automatically {\sl for zero velocity}), and also transitivity holds. Thus `being at rest' is an equivalence relation between admissible cs's , partitioning $\Phi$ into {\sl static equivalence classes}. 

\begin{df} The group of all $4\times 4$ matrices of the form $\Si_S$ with $S\in SO(3)$, where

\be \Si_S = \left(\ba{cc} S  & \0 \\ \0^T & 1 \ea\right), \ee{si}

\n
will be denoted by $SO_4 (3)$. They are called {\bf spatial rotations}. 

The group of all {\bf (space-time) translations} 

\[ T_b : \R^4\ra \R^4, \; T_b(x) = x+b \]

\n
will be denoted by $T(\R^4)$.\end{df}

\n
\textbf{\textsc{Axiom 6}} [{\bf mutual rest}] {\sl Admissible cs's which are mutually at rest differ by a spatial rotation and a space-time translation, i.e.  for any two such cs's $\phi, \phi'$ there exist $S\in SO(3)$ and $b\in \R^4$ such that}

\[ \phi' = T_b\circ \Si_S \circ \phi . \]

Axiom 6 means that mutually at rest, admissible cs's must share the same space distance and, up to a translation, the same time coordinate. In particular they must have the same synchrony relation. This is a standard simplifying requirement, ruling out conventionalist $\ep$-theory (\S 6.2), and should not be confused with its converse, which requires that if $\phi$ is an admissible cs and $S$ is {\sl any} spatial rotation, then $\Si_S \circ \phi$ is also admissible (this is ``spatial isotropy'', our Axiom 12, to be introduced at a later stage). 

The principle of relativity is often considered to have as a consequence, or as its algebraic counterpart, the group property for the set of all transition functions between admissible cs's (e.g. \cite{bll68}).\footnote{In Einstein's article \cite{ei05} (p. 907) one finds a statement to the effect that the special transformations {\sl must} form a group, with no further elaboration or explanation.}   Actually, the essence of ``relativity'', historically speaking, is the requirement that some fundamental physical laws be expressed in the same way in all cs's belonging to a certain, non-static\footnote{Incidentally, for systematic as well as for historical reasons, I think that room should be left in the presentation of the principle of relativity also for `Aristotelian' and `Newtonian' static space-times.} fundamental class $\Phi$. A group $\GC$, or more exactly a subgroup of the transformation group of $\R^4$,  may be involved in the selection of such a class in a natural way, that is, when the class of admissible cs's happens to be the orbit of a single cs under the action of $\GC$ on $\Bi (\EC, \R^4)$. However that a group arises (in this sense) from requiring invariance of a number of laws cannot in general be taken for granted: it depends on the form of the law(s) (cf. \S 6).

I state the existence of a structure group as one of the axioms:

\n
\textbf{\textsc{Axiom 7}} [{\bf structure group}] {\sl There is a subgroup $\GC$ of $\Bi(\R^4)$ such that \( \Phi = \GC\cdot\phi \) where $\phi\in\Phi$ and the ``dot'' at the righthand side denotes the natural left action of $\Bi (\R^4)$ on $\Bi (\EC, \R^4)$ (i. e. composition of maps)}. 

\n
Since $\GC$ is a group, it clearly does not matter which $\phi\in\Phi$ is chosen in Axiom 6. Equivalently, all the maps:

\be F^\phi: \Phi \ra \mbox{Bi} (\R^4), \; \phi'\mapsto \phi'\circ \phi^{-1} \ee{fungroup}

\n
have the same image (i.e. $\GC$) no matter how $\phi\in \Phi$ is chosen. 

From Axioms 6 and 7 it follows that the subset of all $g\in \GC$ having indentically vanishing 3-velocity function is in fact a subgroup of the {\sl Newton group} $\GC_N = SO_4 (3) \rtimes T(\R^4)$ (cf. \cite{mm12}), and the static equivalence classes are its orbits in $\Phi$. For future reference I shall call this subgroup the {\sl rest subgroup}  of $\GC$.

\section{Inertial equivalence and physical velocities (Axioms 8-10)}

Classical and special relativity both assume the validity of the law of inertia, which permits a drastic simplification in the form of the transition functions between cs's. This law, in the present context, can be interpreted as claiming that there is a class of worldlines which are {\sl uniformly moving} in an absolute sense, i. e. they have constant 3-velocities for all admissible cs's. Dynamics adds an explanation why these worldlines are so privileged, but in order for the inertiality requirement to be stated and put to work, no dynamical complements are needed.

There is a classic theorem in the foundations of geometry (see e.g. \cite{jj}), according to which every bijection $F$ from $\R^n$ to itself (with $n>1$) which maps all straightlines into straightlines is an affinity, i. e. it is a map of the form

\[ F: \R^n \ra \R^n, \; x \mapsto Bx + b, \; \mbox{with} \; B\in GL(n,\R),  b\in \R^n .\]

\n
This theorem, however, cannot be used directly in the present setting to conclude to inertiality, contrary to what some authors assume (e.g. \cite{fr12}, p. 750; more weakly, \cite{din}, p. 29, \cite{pau}, p. 9). In fact it is a priori unreasonable, and a posteriori inconsistent, to assume  that {\sl all} straightlines according to some admissible cs are worldlines. This is true neither in Newtonian physics nor in special relativity. In Newtonian physics the straightlines which are not worldlines are those (and only those) contained in any simultaneity space; in special relativity, more embarrassingly, the straightlines which are not worldlines {\sl cannot even be obtained as limits of sequences of worldlines}, since they are, as is well known, all straightlines passing through an event and lying outside the lightcone with vertex at that event.\footnote{Weyl \cite{we52} (pp. 179, 313) sketches a proof which applies a weaker form of the theorem to the case that the inertial worldlines belong to a ``given, arbitrarily thin, cone''. Also Schwartz (\cite{s62}) recognizes the difficulty, but his adaptation of the theorem is not satisfactory, since it assumes, unwarrantedly in its context, that the equation of an hyperplane must be affine.}  

\begin{df} A worldline is {\bf uniform with respect to $\phi\in \Phi$} if its 3-velocity is constant according to $\phi$; it is {\bf uniform}  if it is uniform with respect to every $\phi\in\Phi$.\end{df} 

\n
We need to assume something on how many uniform worldlines exist with respect to every admissible cs. The following is a reasonable assumption:

\n
\textbf{\textsc{Axiom 8}} [{\bf physical velocities}]  {\sl For every admissible cs $\phi$ and $p\in \EC$, the subset  $\hat{\VC}_{\phi, p} \sq \R^3$ of 3-velocities of worldlines through $p$ which are uniform with respect to $\phi$ is a star-shaped neighborhood of $\0$}.

Under this assumption, not all speeds turn out to be physically possible in all directions according to any given $\phi$: it is only assumed that for every direction, worldlines with small enough constant velocity (how small possibly depending on the direction) exist and that in every direction there is a half-interval of speeds starting at $0$. We shall see that this level of generality applies to concrete and not particularly `exotic' examples (\S 6.2).

Notice that if $\phi$ and $\phi'$ are {\sl affinely equivalent}, i. e. $\phi'\circ \phi^{-1}$ is an affinity, then we can write uniquely (cf. \rf{cto} and Axiom 5)

\be x'= Bx + b, \; \mbox{where}\; B = \left(\ba{cc} A & -A\Vy \\ \ky^T & \al \ea\right), \; b\in \R^4, \;  \mbox{and}\; \al >0,  \ee{co_ch}

\n
where $A$ is a $3\times 3$ matrix with $\det A >0$ and $\Vy$ is the (constant) velocity with respect to $\phi$ of any point at rest in $\phi'$. We shall also say that $\Vy$ is the {\sl velocity of the matrix} $B$ and write: $\Vy = \Vy_B$. Moreover

\[ 0< \frac{\p t}{\p t'} = \frac{\det A}{\det B} = \det   \left(\ba{cc} I_3 & -\Vy \\ \ky^T & \al \ea\right)^{-1} = \frac{1}{\al + \ky\cdot \Vy}, \]

\n
from which it follows that 

\be \al + \ky\cdot \Vy > 0. \ee{cto_in}

\begin{df} Two admissible cs's are {\bf inertially equivalent} if all worldlines which are uniform with respect to one are uniform also with respect to the other. A space-time structure is {\bf inertial} if all its members are inertially equivalent. \end{df}

The theorem which contains what is needed about the relationship between inertiality and affinity is the following:

\begin{thm} If $\phi$ and $\phi'$ are inertially equivalent, then they are affinely equivalent, and conversely.\end{thm}  

Proofs of similar statements can be found in the literature (e.g. \cite{f59}), but for completeness' sake a proof fitting the present formalism and assumptions (and in particular Axiom 8) is provided in the Appendix.

\n
\textbf{\textsc{Axiom 9}} [{\bf inertiality}] {\sl All admissible cs's are inertially equivalent, and any cs which differs from an admissible cs by any space-time translation is also admissible}.

From Axioms 7 and 9 and Theorem 3.3, we can conclude that $\GC$ is a subgroup of $Aff (\R^4)$, which is the affine group of $\R^4$.

Remember that a group $G$ is a {\sl semi-direct product} of two subgroups $H$ and $K$ if it is generated by their union, if their intersection is trivial and if one of them is normal; in case $K$ is the normal subgroup we write $G = H\rtimes K$.  The next proposition is a special case of Proposition 2.4 of \cite{mm12}.

\begin{pro} Structure group $\GC$ is of the form

\be \GC = \GC_0  \rtimes T(\R^4) , \ee{gc} 

\n
where $\GC_0$ is a subgroup of the general linear group $GL(4,\R)$. \end{pro}

\n
{\bf Proof}  Given Theorem 3.3 and Axiom 9, we have to show that if 

\[ T(\R^4) \leq \GC\leq Aff(\R^4), \]

\n
then $\GC$ is of the form \rf{gc}: but this follows easily from the fact that \( Aff(\R^4) = GL (4,\R) \rtimes T(\R^4)\). \hfill $\Box$

We now introduce an axiom guaranteeing the `abundance' of admissible cs's and specifying their relationship with physical velocities.

\n
\textbf{\textsc{Axiom 10} [abundance of cs's]}  {\sl There are $\phi\in\Phi$ and $p\in \EC$ such that for every $\Vy$ lying in the interior of $\hat{\VC}_{\phi,p}$, there exists a $\phi' \in \Phi$ such that the velocity of $\phi'$ with respect to $\phi$ is $\Vy$; conversely, for every $\phi'\in \Phi$ the velocity of $\phi'$ with respect to $\phi$ lies in the interior of $\hat{\VC}_{\phi,p}$}.

\n
{\bf Remark} In order to show the need for Axiom 10, one can take as $\GC_0$ the subgroup of the proper orthochronous Lorentz group $\LCU^+$ formed by all matrices having $(0,0,1,0)$ as their 3rd row. Then, clearly, the relative velocity of $\phi'$ with respect to $\phi$, for any $\phi'\in \GC_0\cdot \phi$,  can never attain (or exceed) $c$, but the line represented in $\phi$ by the equations $x^1 = 0, x^2 = 0,  x^3 = tU$, with $U\neq 0$ constant, is a worldline whose velocity in $\phi$ may be as large as desired. Thus Axiom 10 would fail. 

Notice, also, that Axioms 8 and 10 taken together rule out `static' structure groups such as, e.g., the Newton group $\GC_N$. \hfill $\Dia$

\begin{pro} The set $\VC:= \Int (\hat{\VC}_{\phi, p})$ is independent of both $p$ and $\phi$ and starshaped.\end{pro}

\n
{\bf  Proof} The set $\hat{\VC}_{\phi, p}$  does not depend on $p$: in fact if  $\G$ is a uniform worldline through the origin $o = \phi^{-1} (0)$ of $\phi$ and $\phi (p) = b$ , then $\phi^{-1} (\phi (\G) + b)$ is a uniform worldline through $p$, as is easy to verify, and the 3-velocity of the latter at $p$ according to $\phi$ coincide with the 3-velocity of $\G$ at $o$. So $\hat{\VC}_{\phi, p} = \hat{\VC}_{\phi}$, but we also have that $\Int (\hat{\VC}_{\phi})$  does not depend on $\phi$ either, because by Proposition 3.4 and  Axiom 10 

\[ \Int (\hat{\VC}_\phi) = \{\Vy_B \; : \; B\in \GC_0 \}, \]

\n
where the righthand side clearly does not depend on $\phi$ (cf. \rf{fungroup}). 

As to the last point, for every $t\in]0,1[$ the set $t\cdot \VC$ is open and contained in $\hat{\VC}_\phi$, for any $\phi$ (Axiom 8), therefore  

\[ t\cdot\VC \sq \Int (\hat{\VC}_\phi) = \VC, \]

\n
and the thesis follows. \hfill $\Box$

\n
{\bf Remark} The distinction between worldlines having a 3-velocity always in $\VC$ and worldlines with a velocity possibly on the boundary of  $\VC$ is the distinction between (worldlines of) {\sl particles}  (on which an instantaneous admissible cs can be based, at any fixed event) and {\sl signals} (on which, if they are not particles, there are events at which an instantaneous cs cannot be based). Nothing in our argument until and including Theorem 5.1 depends on the assumption that there are signals which are not particles.  \hfill $\Dia$

The following proposition states an intuitively plausible fact about velocities of cs's.

\begin{pro} Let $\phi, \phi_1, \phi_2 \in \Phi$, with $\Phi$ inertial, be such that the velocities of $\phi_1$ and $\phi_2$ with respect to $\phi$ are equal; then $\phi_1$ is at rest with respect to $\phi_2$.\end{pro}

\n
{\bf Proof}  Suppose that $\phi_1$ and $\phi_2$ are admissible cs's, having the same velocity $\Uy$ with respect to $\phi$; we fix our notation by writing 

\[ x_1 = B_1 x +b_1, \; x_2 = B_2 x + b_2, \; \mbox{with}\; \Vy_{B_1} = \Vy_{B_2} = \Uy. \]

\n
Clearly  $x_2 = Bx_1 + b_3$ for a suitable $b_3 \in \R^4$ and $B:= B_2 B_1 ^{-1}$. Since $B_2 = BB_1$, a simple computation gives the velocity $\Vy_{B_2}$ of $\phi_2$ with respect to $\phi$ as

\be \Vy_{B_2} = (I_3 -A_1^{-1} \Vy \ky^T_1)^{-1} (\Uy +\al_1 A_1^{-1}\Vy) . \ee{add_vel}

\n
Equating the LHS to $\Uy$ we get  \( \Uy -(\ky_1\cdot \Uy)  (A_1^{-1} \Vy)  = \Uy +\al_1 A_1^{-1}\Vy\), and therefore \( (\al_1 + \ky_1\cdot \Uy) (A_1^{-1} \Vy)  = \0 \). Now, by \rf{cto_in}, $\Vy$ must vanish, that is, $\phi_1$ and $\phi_2$ are mutually at rest. \hfill $\Box$

\begin{cor} Let $B_1, B_2 \in \GC_0$; if $\Vy_{B_1} = \Vy_{B_2}$, then there is a $S\in SO(3)$ such that $B_2 = \Si_S B_1$. \end{cor}

\n
{\bf Proof}  Take $\phi \in \Phi$ and define $\phi_i = B_i \phi$ ($i=1,2$); then apply Axiom 6 to the previous proposition. \hfill $\Box$

\subsection{Space-time `homogeneity'}

The topic of inertiality has often been discussed in a somewhat misleading way, under the label of ``space-time homogeneity'', following the lead of Einstein's 1905 article.\footnote{``First of all, it is clear that the equations must be {\sl linear} because of the properties of homogeneity which we attribute to space and time'' (\cite{ei05}, p. 898; italics in the original). Cf. \cite{r36}, \cite{c58}, \cite{ll76}.} Let us first explain theoretically how this view has come about (cf. e.g. \cite{l36}, \cite{e67}). 

Any admissible cs defines its own {\sl associated affine structure} on $\EC$, which is obtained by introducing a vector space $V(\EC, \phi)$ of translations in the natural way:

\[ T_{(\phi,b)}: \EC \ra \EC, \; p \mapsto\phi^{-1} (\phi(p) +b), \; b\in \R^4 .\] 

\n
Now let $\phi'$ be another admissible cs, and suppose that 

\be V(\EC, \phi') = V(\EC, \phi).\ee{hom}

\n
This means that for every $b\in \R^4$ there is a $b'\in \R^4$ such that 

\[ T_{(\phi,b)} = T_{(\phi',b')},\]

\n
or equivalently, denoting $F: = \phi'\circ\phi^{-1}$, we must have that for every $b\in \R^4$, $F\circ T_b\circ F^{-1}$ is a translation; that is, $F$ lies in the normalizing subgroup of $T(\R^4)$ in the homeomorphism group of $\R^4$. That this subgroup coincides with the affinity group $Aff(\R^4)$ is easy to verify.\footnote{Details are given, for instance, in \cite{bg69}, p. 1519. Other arguments assume a differentiability condition (cf. \cite{e67}).} 

Condition \rf{hom} means that the affine structures associated to $\phi$ and $\phi'$ are the same, but this is, of course, much more than simply asking for the space-time to be ``homogeneous''. In fact homogeneity is standardly defined as the property of a set-group pair where the group acts on the set transitively; but not all homogeneous spaces are affine, not even those of more evident geometric or physical interest. 

For instance, we can define $M_k$ as what we may call the {\sl generalized Minkowski space-time} $M_k$: \footnote{These space-times are topologically, but not metrically, equivalent to the Friedmann space-times, of well-known cosmological relevance.}

\[ M_k = S_k \times \R , \; k = 0, 1, -1,  \]

\n
where $S_k = \R^3, S^3, H^3$, which are the Euclidean, spherical and hyperbolic space forms with curvature $k = 0, 1, -1$ respectively, $d\si_k^2$ is the standard Riemannian metric on $S_k$, and the space-time metric is:  

\[ ds^2 = d\si_k^2 - c^2 dt^2 .\]

\n
Clearly $M_0$ is Minkowski space-time, but  $M_k$ is a homogeneous space also for $k = \pm 1$ with respect to its own isometry group; nevertheless it is not an affine space. Notice that $M_1$ is the static Einstein space-time of 1917, with unit radius of the universe (\cite{ei17}), and that $M_{-1}$ is also diffeomorphic to $\R^4$. The reason we value particularly the affine group is its link with the inertial motions (Theorem 3.3), not that it is the only way we can conceive of space-time as being ``homogeneous''. Therefore the emphasis on ``space-time homogeneity'' as a supposedly more modest, or clarifying, assumption than inertiality is in fact question-begging.\footnote{The obscurity of the derivation of affinity from homogeneity is pointed out, for instance, in \cite{mi31} and \cite{na32}.  In the present context proofs based on the invariance of d'Alembert equation or on light signalling (\cite{w64}, \cite{whitr}, \cite{g35}) are obviously not relevant.}

\section{Special transformations (Axiom 11)}

The notion of classical coordinate systems related by an affine transformation, with parallel spatial axes and same space-time origins: 

\[ \left\{\ba{rcl} \ry'&=& a (\ry - t\Vy) \\ t' &=& b t \ea\right.,  \]

\n
with $a,b>0$, generalizes to a transformation of type \rf{co_ch} by requiring that: 1) $\Vy$ is a positive eigenvector of $A$, 2) $A$ restricted to $[\Vy]^\perp$ is a positive homothety, and 3) synchrony is preserved (at least) on 2-planes orthogonal to $\Vy$. This leads to the following definition.

\begin{df} A {\bf special transformation} with velocity $\Vy = v\uy$, $|\uy|=1$, is a linear map of $\R^4$ whose matrix is a {\bf special matrix}, that is of the type: 

\be \la \left(\ba{cc} I_3 + (a-1)\uy\uy^T & -a\Vy \\ b\Vy^T & \hal \ea\right). \ee{st1} 

\n
with $a,\hal,\la>0$. A pair  of admissible cs's $(\phi, \phi')$ is a {\bf special pair}  if $\phi'\circ\phi^{-1}$ is a special transformation.\end{df}

The next axiom is implicit in most, if not all, treatments of special relativity which do not start from light-velocity invariance. It remains hidden in standard (1+1)-dimensional treatments because under the assumption of just one spatial dimension {\sl all}  inertial tranformations must be `special'.

\n
\textbf{\textsc{Axiom 11}} [{\bf special transformations}] {\sl There exists a unit vector $\uy\in\R^3$ and a differentiable map defined on an open interval $I\sq \R$ 

\[ B\equiv B_\uy : I \ra \GC, \; \mbox{with}\;  \Vy_{B(v)} = v\uy,\; B(0) = I_4, \]

\n
such that the image of $B$ comprises all special transformations in the direction $\uy$}. 

\n
{\bf Remark} Even when Axiom 11 is satisfied for some $\uy$ in a given theory of relativity, this does not mean that in that theory the special transformations are the `typical'  form of the coordinate changes, to which, that is, reduction is always possible via a spatial rotation. In fact in the structure group there might not be enough spatial rotations. \hfill $\Dia$  

It is easy to check that the product of any two such transformations $B (v_1)B (v_2)$ is a special transformation in the same direction, and therefore it can be written as $B(w)$ for a unique $w$, and the same is true for the inverse of every special matrix. Thus the image of $B$ is a one-dimensional Lie subgroup of $\GC$ diffeomorphic to $\R$, which we denote by $H(\uy)$ and which can be identified with a subgroup of $GL(4,\R)$.  

We denote $w$, as just defined, by $v_1\ast v_2 $, that is\footnote{Of course in \rf{sum} $B^{-1}$ stands for the inverse of the map, not the inverse of the matrix.}

\be v_1\ast v_2 := B^{-1} (B(v_1)B(v_2)). \ee{sum}

Now by computing the product matrix $B(v_1)B(v_2)$ and by defining:

\be f(v): = \hal (v)/a(v), \; g(v): = b(v)/a(v), \ee{fun}

\n
we obtain the following relationships:  

\be \left\{\ba{rcl}  
a(v_1\ast v_2)  &=& a(v_1) a(v_2) (1-g(v_2) v_1 v_2), \\ [5pt]
a(v_1\ast v_2)  v_1\ast v_2 &=&  a(v_1) a(v_2) (f(v_2) v_1 + v_2), \\ [5pt]
b(v_1\ast v_2)  v_1\ast v_2 &=&  a(v_1) a(v_2) (g(v_1)v_1  + f(v_1) g(v_2) v_2), \\ [5pt]
\hal (v_1\ast v_2) &=& a(v_1) a(v_2) (f(v_1) f(v_2) - g(v_1) v_1 v_2), \\ [5pt]
\la (v_1\ast v_2) &=& \la (v_1) \la (v_2). \ea\right.\ee{blo}

Dividing the second equation by the first we get: 

\be v_1 \ast v_2 = \frac{f(v_2) v_1 + v_2}{1 - g(v_2) v_1 v_2}. \ee{vel_add}

\n
Notice that $f$ is always positive and $f(0) =1$. The following statement provides a full characterization of the subgroup of special transformations in a given direction.

\begin{thm} The subgroup $H(\uy)$ of $\GC$ of the special transformations in the direction $\uy$ is given by all matrices of the form

\be  B_\uy (v) = \la (v) \left(\ba{cc} I_3 + (a(v) -1)\uy\uy^T & -a(v)v \uy \\ m a(v) v \uy^T & a(v) (1- lv) \ea\right), \ee{st2}

\n
where $l, m$ are dimensional constants and $m\leq 0$. 

1) If $m = 0$, the domain of $B_\uy$ is $\R$ if $l =0$; and $]-\infty, c_1[$ or $]-c_1, +\infty[$, where $c_1 : = 1/|l|$, if $l\neq 0$. The corresponding functions $a(v)$ and $\la (v)$ are, in the first case ($l =0$):

\be \left\{\ba{rcl} a(v) &=& \dss e^{a_1 v}, 
\\ [4pt]  \la (v) &=& \dss e^{\la_1 v}, 
\ea\right. \ee{form_a_la}

\n
and in the second case ($l\neq 0$):

\be \left\{\ba{rcl} a(v) &=& \dss (1-lv)^{r_1}, 
\\ [4pt]  \la (v) &=& \dss (1-lv)^{r_2}. 
\ea\right. \ee{form_a_lab}

\n
for arbitrary $a_1, \la_1, r_1 , r_2 \in \R$.

2) If $m<0$, let $c_0>0$ such that $m = - 1/c_0^2$. Then for arbitrary $r_1 , r_2 \in \R$ and $\bt = v/c_0$,

\be \left\{\ba{rcl} I &=& ]-c_0 p, c_0 \ol{p}[ \\ [5pt] 
a(v) &=& \dss\left(\frac{1+\ol{p}\bt}{1-p\bt}\right)^{r_1}\frac{1}{\sqrt{1-2\eta\bt-\bt^2}} = 
\frac{(1+\ol{p}\bt)^{r_1 - 1/2}}{(1-p\bt)^{r_1+1/2}}, 
\\ [4pt]  \la (v) &=& \dss \left(\frac{1+\ol{p}\bt}{1-p\bt}\right)^{r_2}, 
\\[4pt]  p : &=& \sqrt{1+ \eta^2} +\eta, \ol{p} := \sqrt{1+ \eta^2} -\eta, \; \eta :=\dss \frac{lc_0}{2}.\ea\right. \ee{form_a_lac}

Moreover, if $\Si_S \in SO_4 (3)$ is an admissible transition function, we have

\be \Si_S B_\uy (v) \Si^T_S = B_{S\uy}(v).\ee{carlo}

\end{thm}

\n
{\bf Proof} Standard Lie group theory implies that $H(\uy)$ is commutative (in fact the only 1-dimensional Lie groups are, up to isomorphism, $S^1$ -- the unit circle group -- and $\R$). The commutativity condition \( B(v_1) B(v_2) = B(v_2) B(v_1)\), is equivalent, given \rf{blo}$_2$ and \rf{blo}$_3$, to the following identities for all $v_1, v_2$:

\[ 1- g(v_2) v_1 v_2 = 1- g(v_1) v_1 v_2, \; v_1 + f(v_1)v_2 = v_2 + f(v_2)v_1. \]

\n
From the first one it follows that $g$ is constant (and we denote by $m$ its value) and from the second one it follows that \( l: = \frac{1 - f(v)}{v}\) (for $v\neq 0$) is another constant. It follows that (cf. \rf{fun}) :

\be f(v) = 1-lv, \; \hal (v) = a(v) (1-lv). \ee{alp}

\n
By suitably exploiting the velocity addition law we shall now obtain stronger conditions on the coefficients. In the new notation, \rf{vel_add} can be re-written as:

\be v_1 \ast v_2 = \frac{v_1 + v_2 - lv_1 v_2}{1 - m v_1 v_2}, \ee{vel_add1}

\n
where $v_1 , v_2$ range over $I$. Since $(I, \ast)$ is a Lie group, the interval $I$ must be invariant under every left translation:

\be h_{v_0} : I \ra I, \; v \mapsto v_0\ast v = \frac{v_0 + (1-lv_0)v}{1-mv_0 v}, \ee{linv}

\n
with $v_0\in I$. Now taking the derivative of  $h_{v_0}$ we obtain:

\[ h_{v_0}'(v) = \frac{mv_0^2- lv_0 + 1}{(1-mv_0 v)^2},  \] 

\n
so the sign of $h_{v_0}'$ is constant (and nonzero), and since $h'_0 \equiv 1>0$, by continuity it follows that  $h_{v_0}$is a strictly increasing function mapping $I$ onto itself for every $v_0\in I$. In particular, the endpoints $w_1, w_2$ must be fixed points for the continuous extension to the `closure' of $I$ (possibly including $\pm\infty$) of every such $h_{v_0}$, i.e. 

\[ \lim_{v\ra w_1} h_{v_0} (v) = w_1, \lim_{v\ra w_2} h_{v_0} (v) = w_2 .\] 

\n
This means that $w_1, w_2$ are either infinite (one of them or both) or (real) solutions of the quadratic equation:

\be mv^2 - lv + 1 =0 .\ee{qe}

\n
Notice that, since $1-lv = f(v)>0$ for all $v\in I$, $m$ is necessarily negative or zero. If $m\neq 0$, then we can introduce a positive constant $c_0$ and write $m = -1/c_0^2$.  

Let $l=0$. If also $m=0$, then it must be $I =\R$ because of the left invariance under \rf{linv}. If $m\neq 0$, then $I = ]-c_0, c_0[$. Of course these conditions give the Galilean and Lorentzian addition laws, respectively. 

1) Let $m=0$ and suppose that $l\neq 0$. According to whether $l>0$ or $l<0$, if $m =0$ then $I = ]-\infty, c_1[$ or $I = ]-c_1, +\infty[$, respectively, where $c_1 = 1/|l|$.

By substituting $g(v) = m=0$ in \rf{fun}$_2$ we obtain immediately $b =0$ and then from \rf{blo}$_1$ and \rf{blo}$_2$, for any $l$, we deduce the functional equations for
$\la$ and $a$:

\be a(v_1 + v_2 - lv_1 v_2) = a(v_1) a(v_2), \ee{eq_a} 

\be \la (v_1 + v_2 - lv_1 v_2) = \la(v_1)\la(v_2). \ee{eq_la} 

By partially differentiating \rf{eq_a} with respect to $v_1$ and then putting $v_1 = 0, v_2 = v$, we obtain the first order differential equation:

\[ \frac{da}{dv} - \left(\frac{a'(0)}{1-lv}\right) a = 0, \]

\n
from which it follows, if $l=0$, \( a(v) = e^{a_1 v} \), and if $l\neq 0$:

\[ a(v) = (1-lv)^{r_1}, \; \mbox{where}\; r_1 : = - a'(0) / l. \]

\n
By the same argument for $\la$ we get, respectively, \( \la (v) = e^{\la_1 v}\) if $l=0$, and 

\[ \la (v) = (1- lv)^{r_2}, \; \mbox{where} \; r_2 : = - \la'(0)/l \]

\n
if $l\neq 0$. 

2) Consider now $m<0$; by solving \rf{qe}  for $m = -1/c_0^2$, we get 

\be I = ]-c_0 p, c_0 \ol{p}[, \; \mbox{where}\; p := \sqrt{1+ \eta^2} +\eta, \ol{p} := \sqrt{1+ \eta^2} -\eta, \eta := \frac{lc_0}{2}. \ee{interv}

\n
Notice that $\eta$ is a dimensionless parameter and that $p\ol{p}=1$; from \rf{alp} we have $\hal (v) = a(v) (1-2\eta\bt)$. By substituting the appropriate values in \rf{blo} we deduce the functional equations for $a$ and $\la$, respectively:

\be a(\frac{v_1 + v_2 - lv_1 v_2}{1 + v_1 v_2/c_0^2}) = a(v_1) a(v_2) (1+\frac{v_1 v_2}{c_0^2}), \ee{eq_b} 

\be \la (\frac{v_1 + v_2 - lv_1 v_2}{1 + v_1 v_2 /c_0^2}) = \la(v_1)\la(v_2). \ee{eq_lb} 

\n
By an argument perfectly similar to that used in 1) we get the first order differential equation:

\[ \frac{da}{dv} - \left(\frac{a'(0) + v/c_0^2}{1-2\eta v/c_0 - v^2 /c_0^2}\right) a =0, \]

\n
from which, after a few straightforward computations, the following explicit formula follows:

\[ a(v) = \left(\frac{1+\ol{p}\bt}{1-p\bt}\right)^{r_1}\frac{1}{\sqrt{1-2\eta\bt-\bt^2}}, \]

\n
where \(\bt: = v/c, \; r_1 := \frac{ca'(0) -\eta}{p+\ol{p}} \).  (The second equality in \rf{form_a_lac}$_2$ follows from the identity:  \( 1-2\eta\bt -\bt^2 = (1-p\bt)(1+\ol{p}\bt) \)). The same procedure applied to \rf{eq_lb} leads us to:

\be \la (v) = \left(\frac{1+\ol{p}\bt}{1-p\bt}\right)^{r_2}, \; \mbox{where}\;  r_2 := \frac{c_0 \la '(0)}{p+\ol{p}}. \ee{eq_lc}

\n
Finally, \rf{carlo}  is a direct consequence of  \rf{st2}. \hfill $\Box$

By applying the results of the previous theorem to \rf{vel_add} we obtain:

\begin{cor} The velocity addition laws corresponding to the different values taken by the constants $m,l$ are:

\[ v_1\ast v_2 =\left\{ \ba{rcl}\dss \frac{v_1 +v_2 -\frac{2\eta}{c_0} v_1 v_2}{1+\frac{v_1 v_2}{c_0^2}} \; &\mbox{for}&\;  m =\dss -\frac{1}{c_0^2}, l = \dss\frac{2\eta}{c_0}, \\ [20pt]
v_1 +v_2 \pm\frac{1}{c_1} v_1 v_2  \; &\mbox{for}&\;  m=0,  l =\dss\mp \frac{1}{c_1} \\ [5pt]
v_1 +v_2 \; &\mbox{for}& \;  m=0, \; l=0.
\ea\right. \]
\hfill $\Box$ \end{cor}

\n
{\bf Remark} From \rf{eq_lb} it turns out that $\la : (I,\ast) \ra (\R^+, \cdot)$ is a group homomorphism, and by composition with logarithm this gives a homomorphism from $(I,\ast)$ to $(\R, +)$. An isomorphism is obtained for every $r\neq 0$  as:

\[ f: (I,\ast) \ra (\R, +), \; v\mapsto r \ln  \frac{1+\ol{p}\bt}{1-p\bt}. \] 

\n
For instance, in the (conformal) Lorentzian case, i.e. for  $\eta =0$, we have (taking $r=1$) $f(v) = \tanh^{-1} (v/c_0)$. Clearly, if one slightly modifies Axiom 10 by requiring the existence of a `one-parameter subgroup' (as standardly defined in Lie theory) of special transformations, the `parameter' should be $s = f(v)$, not the speed itself.  Similarly, in \rf{eq_la} with $l = 1/c_1$, we should use $s = \ln (1-v/c_1)$ etc. \hfill $\Dia$ 

\subsection{Explicit formulae}

Adapting \rf{st2} to the usual form for coordinate transformations we obtain, for the case $m =0$ (or $c_0 = +\infty$):

\be \left\{\ba{rcl} 
\ry'  &=& \dss e^{\la_1 v} (\ry + (e^{a_1 v}(\uy\cdot \ry - vt) - \uy\cdot \ry)\uy)
\\ [10pt] t' &=& \dss e^{a_1 +\la_1} t ,
 \ea\right. \ee{st3a}

\n
if $l=0$, and 

\be \left\{\ba{rcl} 
\ry'  &=& \dss (1-lv)^{r_2} (\ry + ((1-lv)^{r_1}(\uy\cdot \ry - vt) - \uy\cdot \ry)\uy)
\\ [10pt] t' &=& \dss (1-lv)^{1+r_1 + r_2} t ,
 \ea\right. \ee{st3b}
 
 \n
if $l \neq 0$. The transformations \rf{st3a}  (resp. \rf{st3b}) reduce to the Galilean case
 
 \be \left\{\ba{rcl} \ry'  &=& \dss \ry - vt \uy = \ry - t\Vy
\\ [5pt] t' &=& t ,
 \ea\right. \ee{st3bis}
 
 \n
 when $\la_1 = a_1 = 0$ (resp. $l=0$).
  
In case $\uy$ is the unit vector of the $x^1$-axis ($\uy = \ey_1$), as in the usual presentation of the special Lorentz transformations, we obtain what we call the {\sl generalized standard special transformations for $c_0 = \infty$}, respectively for $l=0$: 

\be \left\{\ba{rcl} 
x'^1 &=& \dss e^{(\la_1 + a_1)v}\, (x^1 - vt)
\\[5pt] x'^2 &=& \dss e^{\la_1 v}\, x^2 
\\[5pt] x'^3 &=& \dss e^{\la_1 v}\,  x^3 
\\[5pt] t' &=& \dss e^{(\la_1 + a_1)v}\, t, 
\ea\right. \ee{st3c}

\n
and for $l\neq 0$:

\be \left\{\ba{rcl} 
x'^1 &=& \dss (1-lv)^{r_1 + r_2}\, (x^1 + vt)
\\[5pt] x'^2 &=& \dss (1-lv)^{r_2}\, x^2 
\\[5pt] x'^3 &=& \dss (1-lv)\,  x^3 
\\[5pt] t' &=& \dss (1-lv)^{1+ r_1 + r_2}\, t. 
\ea\right. \ee{st3d}

As for the case $c_0<\infty$ we have

\be \left\{\ba{rcl} 
\ry'  &=& \dss \left(\frac{1+\ol{p}\bt}{1-p\bt}\right)^{r_2} \left(\ry + \left(\left(\frac{1+\ol{p}\bt}{1-p\bt}\right)^{r_1}\frac{\uy\cdot \ry - vt}{\sqrt{1-2\eta \bt -\bt^2}} - \uy\cdot \ry\right)\uy\right)
\\ [10pt] t' &=& \dss \left(\frac{1+\ol{p}\bt}{1-p\bt}\right)^{r_1 + r_2}  \frac{(1-2\eta\bt) t - (v/c_0^2) \uy\cdot \ry}{\sqrt{1-2\eta \bt -\bt^2}},
 \ea\right. \ee{st4}
 
 \n
 which reduces to 

\be \left\{\ba{rcl} 
\ry'  &=& \dss \left(\frac{1+\bt}{1-\bt}\right)^{r_2} \left(\ry + \left(\left(\frac{1+\bt}{1-\bt}\right)^{r_1}\frac{\uy\cdot \ry - vt}{\sqrt{1 -\bt^2}} - \uy\cdot \ry\right)\uy\right)
\\ [6pt] t' &=& \dss \left(\frac{1+\bt}{1-\bt}\right)^{r_1 + r_2}  \frac{t - (v/c_0^2) \uy\cdot \ry}{\sqrt{1-\bt^2}},
 \ea\right. \ee{st4bis}

\n
when $\eta = 0$ (or, equivalently, $l=0$). The {\sl generalized standard special trasformation for $c_0 <\infty$} is:

\be \left\{\ba{rcl} 
x'^1 &=& \dss \left(\frac{1+\ol{p}\bt}{1-p\bt}\right)^{r_1 + r_2}\frac{x^1 - vt}{\sqrt{1-2\eta \bt -\bt^2}}
\\[5pt] x'^2 &=& \dss \left(\frac{1+\ol{p}\bt}{1-p\bt}\right)^{r_2} \, x^2 
\\[5pt] x'^3 &=& \dss \left(\frac{1+\ol{p}\bt}{1-p\bt}\right)^{r_2}\,  x^3 
\\[5pt] t' &=& \dss \left(\frac{1+\ol{p}\bt}{1-p\bt}\right)^{r_1 + r_2}\frac{(1-2\eta\bt) t - vx^1/c_0^2}{\sqrt{1-2\eta\bt-\bt^2}}, \ea\right. \ee{ast}

\n
the inverse transformation being:

\be \left\{\ba{rcl} 
x^1 &=& \dss \left(\frac{1+\ol{p}\bt}{1-p\bt}\right)^{-(r_1 + r_2)}\frac{(1-2\eta\bt) x'^1 + vt'}{\sqrt{1-2\eta \bt -\bt^2}}
\\[5pt] x^2 &=& \dss \left(\frac{1+\ol{p}\bt}{1-p\bt}\right)^{-r_2} \, x'^2 
\\[5pt] x^3 &=& \dss \left(\frac{1+\ol{p}\bt}{1-p\bt}\right)^{-r_2}\,  x'^3 
\\[5pt] t &=& \dss \left(\frac{1+\ol{p}\bt}{1-p\bt}\right)^{-(r_1 + r_2)}\frac{t' + vx'^1/c^2}{\sqrt{1-2\eta\bt-\bt^2}}. 
\ea\right. \ee{ast_inv}

We can conclude that the special transformations, for a fixed direction and module of the velocity, form a one-dimensional connected subgroup depending on 1) a constant $c_0$ (finite or infinite) having the dimension of a velocity; 2) constants having the dimension of the inverse of a velocity ($l, a_1, \la_1$); 3) exponents ($r_1, r_2$). 

\n
{\bf Remark} Formulas \rf{st3a}-\rf{ast_inv} seem not to have appeared in print previously. However, the functional equation approach (applied to $a(v)$ only) was used by Frank and Rothe in their 1911 treatment (with one space dimension), and in particular the expression for $a(v)$ (in particular \rf{form_a_lac}) can be found in their paper, with a different notation (\cite{fr11}, pp. 850-4).

\section{Spatial Isotropy and the Fundamental Theorem (Axiom 12)}

The final axiom needed to arrive at the traditional bifurcation (`either the Poincar\'e or the Galilei group') is:

\n
\textbf{\textsc{ Axiom 12}} [{\bf spatial isotropy}] {\sl The group $\GC_0$ contains $SO_4 (3)$}. 

\n
Equivalently, given Axiom 9, the rest subgroup of $\GC$ coincides with the Newton group. 

Because of \rf{carlo}, Axioms 11 and 12 together imply that a one-subgroup of special transformations exists for {\sl every}  direction. 

Let us list by their labels the axioms introduced so far: 1) {\sl space-time structure}; 2) {\sl topology and differential structure}; 3) {\sl causality}; 4) {\sl time orientation}; 5) {\sl spatial orientation}; 6) {\sl mutual rest}; 7) {\sl structure group}; 8) {\sl physical velocities}; 9) {\sl inertiality}; 10) {\sl abundance of cs's}; 11) {\sl special transformations}; 12) {\sl spatial isotropy}. The axioms that mostly have to do with `relativity' are (the second half of ) 9) and 12), which guarantee, respectively, that by space-time translations and by spatial rotations one stays inside the space-time structure (cf. fn. 5), and 11), which ensures that, as in `Galilean' physics, the space-time structure is not static.  

We remember that all the matrices $B$ in the (proper orthochronous) homogeneous Galileo group $(\GC_G)_0$ can be obtained in the form: 

\be B = \Si_{S_1} B_G (v) \Si_{S_2}, \; \mbox{where}\; B_G (v) = \left(\ba{cc} I_3  & -v \ey_1 \\ \0^T & 1 \ea\right).    \ee{ggg}

\n
for all $v\in \R$, and $S_1, S_2\in SO(3)$, while all the matrices in the proper orthochronous Lorentz group can be written in the form:

\be  \La = \Si_{S_1} \La (v) \Si_{S_2}, \; \mbox{where}\; \La (v) = \La (v\ey_1) =  \left(\ba{cc} I_3 + (\al -1)\ey_1\ey_1^T  & -\al v\ey_1 \\ 
-\frac{\al}{c^2 }v\ey_1^T & \al \ea\right).\ee{lll}

\n
Factorizations \rf{ggg} and \rf{lll} show both the structural kinship between the Galileo and the Poincar\'e groups, and the similar role played as building blocks by standard special transformations and spatial rotations. We can now state and prove: 

\begin{thm} [{\bf Fundamental Theorem of the Theory of Relativity}] Let $\Phi$ be a space-time structure satisfying Axioms 1-12. Then the structure group $\GC$ of $\Phi$ is either the Galileo group or the proper orthochronous Poincar\'e group.\end{thm}

\n
{\bf Proof} Let $\phi, \phi' \in\Phi$ be related by an affinity (Axiom 9) whose linear part is given by the  matrix $B$,  with notation like in \rf{co_ch} (in particular $\Vy_B \equiv \Vy$).  Because of Proposition 3.4 we can limit ourselves to dealing with $\GC_0$, that is, with matrices. By Axiom 12, for every $S\in SO(3)$ the map $\Si_S\circ \phi$ belongs to $\Phi$; notice that the transition function from $\Si_{S} \circ \phi$ to $\phi'$ is, up to a translation, $B\Si_{S^T}$, whose velocity is $S\Vy$. It follows from Axiom 10 and Proposition 3.5 that $\VC$ is $SO(3)$-invariant and star-shaped. 

\n
Moreover, if $B_\uy (v)$ is a subgroup of special transformations contained in $\GC$ (Axiom 11), then $\GC$ must contain as well all matrices of the form 
$\Si_{S_1} B_\uy (v) \Si_{S_2}$ for $S_1, S_2\in SO(3)$; in particular,  for a suitable choice of $S_1 = S_2^T = S$, we have, by \rf{carlo}, that also the standard special transformations $B_{\ey_1} (v)$ are in $\GC$. 

\n
Now,\footnote{What follows generalizes Poincar\'e's argument in \cite{poi06}.} if we take the following element of $SO(3)$ 

\[ S_0 = \left (\ba{ccc} -1 & 0 & 0 \\ 0 & -1 & 0 \\ 0 & 0 & 1 \ea\right), \]  

\n
we see that the velocity of $\Si_{S_0}^T B_{\ey_1} (v)  \Si_{S_0}$ is $-\Vy = (-v, 0, 0)$. An easy examination of the explicit formulas for the standard special transformations in the case $m=0$ shows that 

\be \Si_{S_0}^T B_{\ey_1} (v)  \Si_{S_0} = B_{\ey_1} (-v) \ee{inv_poi} 

\n
applies to \rf{st3c} (resp. \rf{st3d}) if and only if $a_1 = \la_1 = 0$ (resp. $ l =0$); while \rf{inv_poi} applies to \rf{ast} if and only if $\eta = r_1 = r_2 = 0$. So we have, in the first case, $B_{\ey_1} (v) = B_G (v)$, and in the second case, $B_{\ey_1} (v) = \La (v)$. Because of the characterizations \rf{ggg} and \rf{lll}, we have that $\GC$ contains, respectively, $\GC_G$ and $\PCU^+$. 

\n
In order to show that in both cases not just inclusion but equality holds, we notice that the only star-shaped open neighborhoods of $\0$ which are $SO(3)$-invariant are I)  $\R^3$ and II) the open balls of finite radius $c_0$ centered in $\0$ (for every $c_0 >0$). Thus in both cases all velocities in $\VC$ are represented by matrices in $\GC_G$ and in $\PCU^+$, respectively, and 
 Corollary 3.7 forbids any other matrices to be contained in the structure group.\hfill $\Box$ 

\n
{\bf Remark} 1. In the proof of  Theorem 5.1, the case in which $c_0$ is a possible physical speed and the case where it is just a supremum need not be separated. In fact at this foundational level the hypothesis that there exist actual signals with speed $c_0$ does not make any difference. In particular the theorem, in itself, is perfectly compatible with whatever assumption on this issue.  \hfill $\Dia$ 

\n
2. An alternative route to the Galilei and Poincar\'e groups passes through the so-called reciprocity principle, which many authors  (e.g. \cite{s36}, \cite{whit}, \cite{p21}, \cite{s52},  \cite{w65}, \cite{d66}, \cite{m66}, \cite{mo72}, \cite{a68}, \cite{pal}) introduced in order to obtain a further simplification after a transformation of type \rf{st1} had been somehow (often not very transparently) arrived at. 

\n
Let us call {\sl reciprocal velocity} of the matrix $B$ the velocity of $B^{-1}$, that is the velocity of $\phi$ with respect to $\phi'$; we denote it by $\Wy \equiv\Wy_B = \Vy_{B^{-1}}$.
The {\sl reciprocity principle} is the condition that the module of the velocity of $\phi'$ with respect to $\phi$ be the same as the module of the reciprocal velocity: $|\Vy| = |\Wy|$. 

\n
This requirement reduces to $\Wy = -\Vy$ {\sl for special pairs} $(\phi, \phi')$; however, for a general inertial transformation (that is for arbitrary pairs linked, say, by a general Galilei or Lorentz transformation) the latter version of  `reciprocity' does not hold, contrary to what seems to be assumed in some treatments (see e.g. \cite{zh97}, p. 18; cf. \cite{mm11}, \S 5.1). 

\n
Now for special pairs we have:

\be \Wy = -\frac{a(\Vy)}{\al (\Vy)}\Vy,  \ee{rec}

\n
so (RP) reduces to the identity \(a = \al\). Thus \rf{st2} obeys the reciprocity principle if and only if $l = 0$ (notice, however, that also \rf{st3a} satisfies it). Under this condition in \rf{ast} we have $p = \ol{p} = 1$, and if we put $r_1 = 0, r_2 = s$ we obtain the transformation: 

\be \left\{\ba{rcl} x'^1 &=& \dss \left(\frac{1+\bt}{1-\bt}\right)^{s}\, \frac{x^1 - vt}{\sqrt{1-\bt^2}}
\\[5pt] x'^2 &=& \dss \left(\frac{1+\bt}{1-\bt}\right)^s \, x^2 
\\[5pt] x'^3 &=& \dss \left(\frac{1+\bt}{1-\bt}\right)^s\,  x^3 
\\[5pt] t' &=& \dss \left(\frac{1+\bt}{1-\bt}\right)^s \, \frac{t- vx^1/c^2}{\sqrt{1-\bt^2}}. \ea\right. \ee{ast_rp}

\n
This one-dimensional subgroup (for every fixed $s \in \R$) is contained in the conformal Poincar\'e group. It has a common null eigenvector and it can be shown to be contained in a 8-dimensional subgroup $\GC$ of the conformal Poincar\'e group with this property. Group $\GC$  has been the object of a series of articles by Bogoslovsky, starting in 1977 (\cite{b}). The Lie algebra of its linear factor (denoted by $\GC_0$ in \cite{mm11}), is generated by a basis $(\si_0, \si_1, \si_2, \si_3)$ verifying the identities:

\[ [\si_0,\si_1] = 0\; [\si_0 , \si_2] = \si_3, \;  [\si_0,\si_3] = -\si_2, \; [\si_2, \si_3] =0, \; [\si_1, \si_2] = -\si_2, \; [\si_1, \si_3] = -\si_3 . \]

\n
Here $\si_0$ corresponds to a 1-parameter subgroup of spatial rotations and $\si_1$ to the only one-dimensional subgroup of special Lorentz transformations. It follows that $\GC$ has a 7-dimensional subgroup (where the Lie algebra of its linear factor is generated by  $(\si_0, \si_2, \si_3)$), which violates both Axioms 10 and Axiom 11. In particular, this subgroup is an example of a non-static structure group which does not admit a one-dimensional subgroup of special transformations. 

\n
A similar kind of anisotropy can be introduced in the classical setting, thus defining an anisotropic Galilei group. A unified approach and a discussion of the properties mentioned here, and others, of these anisotropic groups is contained in \cite{mm11}. 

\section{Two-way light isotropy (Axiom $6^\ast$)}

The class of theories in which both spatial isotropy and reciprocity fail is particularly interesting from the viewpoint of the conventionality of simultaneity issue (\cite{mm01}). Let us introduce a new axiom, which is also the first one which endowes the limiting, direction-dependent speeds, with a concrete physical meaning, that of being the speeds of physical signals (``light'').

\n
\textbf{\textsc{Axiom $6^\ast$}} {\sl The two-way isotropy of the speed of light holds, with a constant $c$}.

The two-way isotropy of the speed of light may be concisely expressed by the equation (cf. \S 2 in \cite{mm01}) :

\be \frac{1}{c_+} + \frac{1}{c_-} = \frac{2}{c}. \ee{tw}
 
 Conventionalists hold that only the round-trip, or two-way, isotropy of light is a law with an unambiguous physical content, and that the standard one-way version of the law adds nothing to the former except for the choice of a convention -- one among infinitely many, justified essentially by formal expediency (\cite{mm01},\cite{mm12}). Now, from the conventionalist viewpoint {\sl also the assumption that the transition functions among admissible cs's form a group (Axiom 7) must be classified as a convention} of the same kind. In fact if  we adopt the round-trip law (instead of the `classical' one-way law) as our theory's cornerstone, the space-time structure $\Phi$ does not come out naturally as a group orbit, but as a {\sl union} of infinitely many group orbits with respect to different subgroups of $\Bi (\R^4)$, as we are going to see. 

We shall now derive and briefly discuss the transformation between inertial cs's in which the round-trip isotropy of light is verified.  The derivation has been done with varying degrees of generality and explicitness by several authors (e.g. \cite{zh97}, \cite{se96}, \cite{avs98}, \cite{se99}). The following derivation strikes me as both simpler and more logically transparent than others. 

Recall that an {\sl inertial basis} in $\R^4$ with the usual Lorentz structure:

\[ g_c : \R^4 \times \R^4 \ra \R, \; (x,y)\mapsto x^1 y^1 + x^2 y^2 + x^3 y^3 -c^2 x^4 y^4 \equiv x^T G_c y, \]

\n
 is a basis $b = (w_1 , w_2 , w_3, w_4)$ where $w_1, w_2, w_3$ are space-like and $w_4 \equiv u$ is a future timelike vector (cf. \cite{mm01}, p. 788). An {\sl inertial cs} is an affine coordinate system whose associated basis is inertial. Let us start reformulating, for ease of reference, Proposition 3.13 of \cite{mm12}:

\begin{pro} In Minkowski space-time, for every inertial cs $\phi$ such that the isotropy of the two-way velocity of light holds, there exists a unique Minkowski coordinate system $\ol{\phi}$ such that the transition function from $\ol{\phi}$ to $\phi$ is of the form:

\be x= \la K\ol{x} \; \mbox{with}\; \la >0,\; K = \left(\ba{cc} I_3 & \0 \\ \ky^T & 1\ea\right),\; |\ky |<1/c . \ee{rei}

\n
Vice versa every $\phi$ related to a Minkowskian $\ol{\phi}$ by \rf{rei} satisfies the two-way isotropy of the velocity of light.  \hfill $\Box$ \end{pro}

Let $\phi$ and $\phi'$ be inertial cs's for which the isotropy of the two-way speed of light holds; we want to find the transition function $\phi'\circ\phi^{-1}$.

With reference to the previous proposition, we have for $\la_1, \la_2 > 0$,  \( x = \la_1 K_1 \ol{x}, \; x' = \la_2 K_2 \ol{x}' \), with $K_1, K_2$  like in \rf{rei}, where 
$\ol{\phi}, \ol{\phi}'$ are Minkowskian cs's systems. Let $\La$ be the Lorentzian matrix giving the linear part of the transition function from $\ol{\phi}$ to $\ol{\phi}'$. It follows that

\be x' = \la B x + b,\; \mbox{where}\; B = K_2 \La K_1^{-1} \ee{rtt}

\n
where \( b= (\by, b^4) \in \R^4\). Taking account of the form of the Lorentz matrices:

\be \La = \left(\ba{cc} A & -A\Vy \\ -\frac{\al}{c^2}\Vy^T & \al\ea\right ), \; \mbox{with}\; A^TA = I_3 + \frac{\al^2}{c^2} \Vy\Vy^T, \; \al = (1-\bt^2)^{-1/2}, \ee{lormat}

\n
where $\Vy$ is the velocity of $\ol{\phi}'$ with respect to $\ol{\phi}$, we have (with $\la = \la_2/ \la_1$):

\[ B = \la\left(\ba{cc} I_3 & \0 \\ \ky_2^T & 1\ea\right)
\left(\ba{cc} A & -A\Vy \\ -\frac{\al}{c^2}\Vy^T A & \al \ea\right)
\left(\ba{cc} I & \0 \\ -\ky_1 & 1\ea\right).  \]

A straighforward computation leads to the following:

\begin{pro} The coordinate change between two cs's obeying Axioms 1-5 and $6^\ast$ is

\be\left\{\ba{rcl} 
\ry' &=&  \la A((I_3 +\Vy \ky_1^T )\ry - t\Vy) +\by \\  [6pt]
t' &=& \la ((A^T \ky_2 -\frac{\al}{c^2}\Vy - (\al -\ky_2^T A \Vy)\ky_1)\cdot\ry + (\al -\ky_2^T A\Vy)t) + b^4 ,\ea\right.
\ee{nsco} 

\n
with $\la >0, b\in \R^4, \Vy\in B(\0,c), \ky_1, \ky_2 \in B(\0, 1/c)$ and $A$ satisfying \rf{lormat}. System \rf{nsco} defines a set $\FC$ of transformations depending on 17 independent parameters. \hfill $\Box$ \end{pro} 

The set $\FC$ of all transformations preserving equation \rf{tw} is not a group (under ordinary map composition), although the inverse of every transformation in $\FC$ is still in $\FC$. This shows that loosely framed claims such as ``{\sl the set of all one-one transformations leaving [unchanged] any set of equations forms a group}'' must be handled with some care.\footnote{The statement, italicized in the original, is quoted from \cite{bir}, p. 100.}  So this gives an example of a theory of relativity (Axioms 1-5) where the basic physical law which is required to hold in all admissible cs's {\sl does not produce in a natural way a structure group for the theory}.

By putting in \rf{nsco}:

\[ \Vy = (v,0,0),\; \ky_1 = (a_1,0,0),\; \ky_2 = (a_2,0,0), \]

\n
and by requiring furthermore that $A$ be diagonal and with positive entries, it follows by \rf{lormat} that $A = \diag(\al,1,1)$, and therefore: 

\be\left\{\ba{rcl} x'^1 &=& \la \al ((1+a_1 v)x^1 - vt)
\\  [4pt] x'^2 &=& \la x^2
\\ [4pt] x'^3 &=& \la x^3
\\ [4pt] t' &=& \la \al ((a_2 - v/c^2 -(1 - a_2 v) a_1)x^1 + (1 -a_2 v)t).\ea\right. \ee{spetra} 

\n
The values for the free parameters which give the special Lorentz transformations are, clearly, $a_1 = a_2 =0$, $\la =1$. These special transformations satisfy Axiom 6 if and only if $a_1 = a_2, \la =1$ ($v=0$ implies $\al =1$), so in general even this axiom is {\sl not} satisfied. 

\n
{\bf Remark} If we renounce Axiom 9, in Proposition 6.1 we can substitute the linear term $\ky\cdot \ry$ with any function. More precisely, for a fixed Minkowski coordinate system $\ol{\phi}$ any coordinate transformation 

\be \ry = \la \ol{\ry},\; t = \la (  \ol{t} + g(\ol{\ry})), \ee{rei0}    
 
\n
where $g : \R^3 \ra \R$ is an arbitrary differentiable function such that $|\na g|<1/c$ (for instance take $g (\ol{\ry}) := (\sin\ol{x}^1)/2c$), defines a coordinate system $\phi$ which satisfies the two-way light isotropy.  This is easily proved (cf. \cite{avs98}, pp. 131-2). Suppose that $\G$ is any lightlike curve and its projection $C$ in the 3-space of $\ol{\phi}$ is closed; then, if $\ol{\ry}: I \ra \R^3$ represents $C$ in $\ol{\phi}$, we have $|\ol{\vy}|\equiv c$ and therefore 

\[\dss\oint_C \frac{ds}{|\vy|} = \frac{1}{c}\dss\oint_C (1+ \na g\cdot\ol{\vy}) ds = \frac{1}{c}\dss\oint_C ds +\dss\oint \na g\cdot d\sy 
= \dss\frac{\ell(C)}{c},\] 

\n
where $\ell (C)$ is the length of $C$.

\subsection{Absolute simultaneity}

The {\sl absolute simultaneity condition} (cf. \cite{mm01}, \S 9)  is the requirement of proportionality between the time coordinates, i.e. $t' \propto t$; for \rf{nsco} this is equivalent to: 

\be A^T \ky_2 -\frac{\al}{c^2}\Vy = (\al -\ky_2^T A\Vy)\ky_1 , \ee{asc}  

\n
and the corresponding sub-family of \rf{nsco} becomes:

\[\left\{\ba{rcl} \ry' &=& \la A((I+\Vy\ky_1^T)\ry -t\Vy) + \by \\ t' &=& \la (\al -\ky_2^T A\Vy) t + b^4. \ea\right.\]

If we scalarly multiply both sides of \rf{asc} by $\Vy$, we get, after a few passages:

\[ \ky_2^T A\Vy =\al\frac{\bt^2 +\ky_1\cdot\Vy}{1+\ky_1\cdot\Vy}, \]

\n
and by substituting this into the time coordinate equation we get:

\be\left\{\ba{rcl} \ry' &=& \la A((I+\Vy\ky_1^T)\ry -t\Vy ) + \by \\ t' &=& \dss\frac{\la t}{\al (1+\ky_1\cdot\Vy)} + b^4, \ea\right.\ee{astor}

\n
which is a set of transformations -- not a group, again -- depending on 14 parameters. In case $\phi=\ol{\phi}$, we have $\ky_1 =\0$ and therefore

\be\left\{\ba{rcl} \ry' &=& \la A(\ry -t\Vy) + \by \\ t' &=& \dss\frac{\la t}{\al} + b^4. \ea\right. , \; \mbox{with}\; 
A^T A = I_3 +\frac{\al}{c^2}\Vy\Vy^T ,\ee{ivest}

\n
which is {\sl the most general transformation from a Minkowskian cs to an inertial cs satisfying both the round-trip isotropy and the absolute simultaneity conditions}. The ``special'' version of \rf{ivest} is:

\be\left\{\ba{rcl} x'^1 &=& \la \al (x^1 - vt),\\ x'^2 &=& \la x^2,\\ x'^3 
&=& \la x^3,\\ t' &=& \dss\frac{\la t}{\al}.\ea\right. \ee{ivests} 

Notice that while $|v|<c$, the corresponding speed of light in a $\phi'$ can be as big as desired.

This set of transformations was discussed by several authors, particularly by Tan\-gher\-lini (\cite{ta61}, (1.16), p. 9); but it is surely the three-part article by Mansouri and Sexl (\cite{ms77a}) which must be credited for its more recent revival (cf. \cite{mm01}, \S9). 

\subsection{Two-way light isotropy and the group axiom}

We have seen that Axiom $6^*$ potentially conflicts with Axiom 7. Nevertheless it makes sense to inquire whether one can enforce Axiom $6^\ast$ along with Axiom 7, that is whether there exist  nontrivial invariance groups for the law of {\sl two-way} constancy of the limiting velocity (`nontrivial' means here violating Axiom 12). In fact there are infinitely many such groups, as we are going to see in a moment. 

\n
Suppose $\GC$ is a structure group of a space-time $(\EC, \Phi)$ satisfying Axiom 9, so that in particular $\GC = \GC_0 \rtimes T(\R^4)$, and every $B\in \GC_0$ is of the form 
\rf{rtt}. Let $\phi\in\Phi$ be fixed, so that $\Phi = \GC\cdot\phi$; let us pick any other $\phi_1\in\Phi$. If $\Phi$ satisfies Axiom $6^\ast$ with respect to a fixed Minkowski structure $\ol{\Phi}$, we have from \rf{rtt} that 

\[ B:= (\phi_1\circ\phi^{-1})_0 =\la  K_1\La K^{-1}, \]

\n
 where $K$ can be considered as fixed, while $\la>0$, $\La\in \LCU^+$ and $K_1$ all depend on $\phi_1$.  A simple computation shows that

\be \Vy_B = \frac{\Vy}{1-\ky\cdot\Vy}, \ee{VVV} 

\n
which shows that Axiom 10 is satisfied if all $\Vy\in B(\0,c)$ are allowed in the Lorentzian factor. Suppose the conformal factor $\la$ is set equal to 1. The following theorem holds. 

\begin{thm} Let $\GC_0$ be a subgroup of $GL(4, \R)$ with all elements of the form \( K' \La K^{-1}\) with $K'$ varying in some nonempty set and $K$ fixed, both $K, K'$ of the form \rf{rei}, and suppose that $\La$ varies in a subset $\XC \sq \LCU^+$ such that

\[ \XC \ra B(\0,c), \; \La \mapsto \Vy_\La \]

\n
is onto. Then $\XC$ is a subgroup of $\LCU^+$ and \( \GC_0 = K \XC K^{-1}\).\end{thm} 

\n
{\bf Proof} It is easy to verify that the set $\XC$ in the statement must be a subgroup of $\LCU^+$. Consider now any product of the form: 

\[( K_2\La_2 K^{-1}) ( K_1\La_1 K^{-1}) =  K_3\La_3 K^{-1}, \]

\n
or, equivalently:

\[ K_4 \La_2 K_5 = \La \in \LCU^+, \; \mbox{with}\; K_4 := K_3^{-1}K_2, \; K_5 := K^{-1}K_1 , \; B:= \La_3 \La_1^{-1}\] 

\n
where $K_4, K_5$ have the same form as in \rf{rei}, except, possibly, for the condition on the upper bound on $|\ky|$. Our claim is that $K_5 = I_4$, or, equivalently, $\ky_5 = \0$ (that is, $\ky_1 = \ky$). A simple computation, taking into account that $\La$ and $\La_2$ are of the form \rf{lormat} gives:

\[ \La  =  \left(\ba{rcl} A_2 (I_3 - \Vy_2 \ky_5^T)  & -A_2\Vy_2 \\ [4pt]  (A_2^T\ky_4 -\frac{\al_2}{c^2}\Vy_2 + (\al_2 -\ky_4\cdot A_2\Vy_2)\ky_5)^T  & \al_2 -\ky_4\cdot A_2\Vy_2\ea\right), \]

therefore 

\be \Vy \equiv \Vy_\La = \frac{\Vy_2}{1-\ky_5\cdot\Vy_2}. \ee{VVV1}

\n
Since both $\La$ and $\La_2$ must satisfy the identities in \rf{lormat}, we have:

\[ \al_2^2 \left(\frac{1}{c^2} \Vy_2\Vy_2^T - (\Vy_2\ky_5^T + \ky_5\Vy_2^T\right ) + V_2^2 \ky_5\ky_5^T) = \frac{1}{c^2}\left(\frac{\al_2 - \ky_4\cdot \Vy_2}{1 - \ky_5\cdot\Vy_2}\right)^2\Vy_2\Vy_2^T. \]

\n
We can choose $\Vy_2$ arbitrarily in $B(\0, c)$, so let us take it as a nonzero vector orthogonal to $\ky_5$, and multiply scalarly both sides by $\ky_5$; we obtain:

\[ |\Vy_2|^2 |\ky_5|^2 \ky_5 = |\ky_5|^2 \Vy_2, \]

\n
and since $\Vy_2 \perp \ky_5$, it must be $\ky_5 = \0$ as required. \hfill $\Box$

One may call any maximal group with the linear factor of the form described in Theorem 6.3 a {\sl Reichenbach group} if $\ky \neq \0$:    

\be \GC_R (\ky): = K\LCU^+ K^{-1} \rtimes T(\R^4) \ee{reig} 

\n
The corresponding Reichenbach function is (cf. \cite{mm01}, p. 791):

\[ \ep (\ry) = \frac{1}{2} (1 + c\ky\cdot \frac{\ry}{r}). \]

In terms of the Reichenbach groups we can express the spacetime structure $\Phi$ as

\be\Phi = \bigcup_{\ky\in B(\0, 1/c)} \R^+\cdot \GC_R (\ky)\cdot K\ol{\phi} \ee{cup}

\n
with $\ol{\phi}$ any fixed cs in the Minkowski structure. 

The linear factor $ \GC_R (\ky)_0$ is the invariance group of the Lorentzian matrix:

\[ G = (K^{-1})^T G_c (K^{-1}) = \left( \ba{cc} I_3 -c^2 \ky  \ky^T & c^2\ky \\ c^2 \ky^T & -c^2 \ea\right),  \]  

\n
and the corresponding space-time metric is:

\[ ds^2 = d\ry^T  (I_3 -c^2 \ky  \ky^T )  d\ry + 2c^2 d\ry\cdot \ky dt - c^2 dt^2 . \]

\n
Therefore the set $\VC$ of relative velocities of admissible cs's -- i.e. those cs's belonging to a fixed class $\Phi_{\ky}$ having $\GC_R (\ky)$ as structure group  (cf. Proposition 3.5) -- is 

\[ \left(\ba{cc} \vy \\ 1 \ea\right)^T G \left(\ba{cc} \vy \\ 1 \ea\right) = \vy^T (I_3 -c^2 \ky  \ky^T )  \vy + 2 c^2\ky\cdot\vy - c^2  < 0, \]

\n
and its boundary is what may be called the {\sl velocity ellipsoid} of $\GC_R (\ky)$:

\be |\vy|^2 - c^2 (1-\ky \cdot\vy)^2 = 0 .\ee{ver}

In order to study this surface suppose $\ky = k \ey_1$. The equation becomes

\[ (v^1)^2 + (v^2)^2 + (v^3)^2 -c^2 (1-k v^1 )^2 = 0, \]   

\n
which is an ellipsoid with a 1-parameter group of rotations around the bigger semi-axis $c/(1-c^2 k^2)$, the other two semi-axes being equal to $c/\sqrt{1-c^2 k ^2}$. Since the centre of the ellipsoid is $(-c^2 k/ (1-c^2 k^2), 0, 0)$, the interval of the speeds in the $x^1$-direction is asymmetric with respect to 0:

\[ c]-\frac{1}{1-ck}, \frac{1}{1+ck}[, \] 

\n 
while in both the $v^2$ and the $v^3$-directions (i. e. in the directions orthogonal to $\ky$) it is the standard $]-c,c[$. 

 The Reichenbach groups form a family of Lie groups parametrized by $\ky$, as $\ky$ varies in the open set $0<|\ky|<1/c$ in $\R^3$. They are all isomorphic to one another, and they are all isomorphic to the Poincar\'e group; explicitly: \( \GC_R (\ky) = K \PCU^+ K^{-1}\). The Reichenbach groups satisfy Axiom 11 for the direction $\uy = \vers (\ky)$. In fact we have:
 
 \[ K \La (v\uy)K^{-1} =  \left( \ba{cc} I_3 + (\al (1+kv)-1) \uy\uy^T & -\al v \uy \\ -\al (\frac{1}{c^2} - k^2)v\uy^T & \al (1- kv) \ea\right), \] 
 
 \n
 which verifies definition 4.1. Notice however that Axiom 6 is not satisfied: in fact for every $S\in SO(3)$ matrix \( B := K\Si_S K^{-1}\) has zero velocity ($\Vy_B = \0$), since 

\[  B = \left( \ba{cc} S & \0 \\ \ky^T (S-I_3) & 1 \ea\right), \]

\n
therefore Axiom 6 would be satisfied if and only if $\ky = \0$, which would take us back to the Poincar\'e group. Equivalently, the rest subgroup of $\GC_R (\ky)$ is $K\GC_N K^{-1}$, which is not contanined in $\GC_N$ unless $\ky = \0$.

Physically, to take a Reichenbach group for some $\ky$ as the structure group of physics corresponds to having a relativistic physics in which the privileged cs's $\phi$ have been all synchronized by choosing the same Reichenbach function $\ep$ with respect to a Minkowskian cs depending on $\phi$. As a consequence {\sl none} of the elements of $\Phi_\ky$ is Minkowskian: indeed, there is a unique Minkowskian structure $\ol{\Phi}$ related to $\Phi_\ky$ by

\[ \ol{\Phi} = K^{-1} \Phi_\ky\]

\n
for a suitable $K$. It must be stressed, however, that this construction goes against the spirit of the Reichenbach's approach, since his $\ep$-argument was advanced only to show that  every single inertial notion of rest (denoted as $\G (u)$ in \cite{mm01, mm12}) could arbitrarily choose its own synchrony with considerable latitude, {\sl irrespective of what other notions of rest did}. As we have seen (cf. \rf{cup}), the genuine mathematical counterpart of Reichenbach's view is to take the space-time structure $\Phi$ as a union of orbits of infinitely many groups. The main interest of a theory based on $\GC_R (\ky)$ is that it is {\sl a theory of relativity satisfying the group axiom and where the law of light propagation holds only in the two-way version}. We can re-state the conventionalist position concerning relativity as holding that there is no physical fact of the matter which allows us to distinguish between any of the Reichenbach groups and the Poincar\'e group.  I have explained in \cite{mm01} and \cite{mm12} why I think this claim cannot be maintained except in a rather uninteresting sense. 

\section {Appendix -- Proof of Theorem 3.3}

It is easy to verify that if $\phi, \phi'$ are affinely equivalent cs's, then they are also inertially equivalent. 

Let us consider the converse. Suppose that every uniform worldline for $\phi$ is also uniform for $\phi'$. Since points at rest for $\phi$ are physical worldlines (Axiom 3),  we have for every $\ry_0 \in \R^3$:

\be \frac{\p\ry'}{\p t}(\ry_0 , t)=\frac{\p t'}{\p t} (\ry_0, t)\Wy  \ee{laprima}

\n 
with $\Wy$ depending at most on $\ry_0$. It follows that the transition function from $\phi$ to $\phi'$ can be re-written as

\[ \left\{\ba{rcl} \ry' &=& t' (\ry , t)\Wy (\ry) + \Uy (\ry) \\ [4pt] t'  &=& t' (\ry , t). \ea\right. \]

\n
where $\Uy: \R^3 \ra \R^3$ is a suitable differentiable function.

{\sl Claim 1}: $\Wy$ is constant. 

In fact let \( \ry (t) = \ry_0 + (t-t_0)\vy\) be any uniform motion for $\phi$ with $\vy$ physical and $\R$ as its domain. If we put  \( f(t):= t' (\ry_0 + (t-t_0 )\vy , t) \) it is easy to see that $f$ is a diffeomorphism of $\R$ onto itself. (Let $\vy'$  the constant velocity of the same worldline according to $\phi'$, and suppose that the interval $f(\R)$ had, say, a finite supremum $b$; then, if $\phi'\circ\phi^{-1} (\ry_0, t_0) = (\ry'_0, t'_0)$ the subset

\[ \phi\circ\phi'^{-1}(\{(\ry'_0 + (t'- t'_0)\vy')\; : \: t'\in [t'_0, b]\}) \]

\n
would have to be noncompact, which is absurd.) Now

\be \vy' = \Wy + \frac{1}{\dof}(f(\vy\cdot\na)\Wy + (\vy\cdot\na)\Uy) \ee{lasec}

\n
must not depend on $t$. Putting $t=t_0$ we get that there is a function $\vy' = \Ky (\ry_0, \vy)$ such that $\Ky (\ry (t), \vy) \equiv \Ky (\ry_0, \vy)$, and 

\[ \dof (t_0)\ (\Ky (\ry_0, \vy) -\Wy (\ry_0)) - f(t_0) ((\vy\cdot\na)\Wy)_{\ry_0} = ((\vy\cdot \na)\Uy)_{\ry_0}. \]

\n
For every fixed $\ry_0$ and physical velocity $\vy$ this is an ordinary differential equation in $f(t)$ (after notation change from $t_0$ to $t$) with vector coefficients:

\[ \dof  \Ay +f \By = \Cy, \; \mbox{where}\; \Ay := \Ky - \Wy, \; \By := -(\vy\cdot\na)\Wy, \; \Cy:= (\vy\cdot\na)\Uy. \]  

\n
Now, if  for all $\ry_0$ we have $\By = \0$ when we choose $\vy$ in three linearly independent directions, then $\na\Wy =0$ and therefore $\Wy$ is constant. But for $\By$ there is no other possible value. Suppose by contradiction, that $\By \neq \0$ for some $\ry_0$ and $\vy$. Then $f$ must be a solution of a (scalar) differential equation with constant coefficients of the form $a\dof + bf = c$ with $b\neq 0$. Now $a\neq 0$, since otherwise $f$ would be constant; it follows that $f$ is of the type \( f(t) = ke^{-bt/a} + c/b \), which clearly is not onto $\R$, no matter what the values of $a,b,c$ are. So claim 1 is proven.

It follows that \( \ry' (\ry ,t) = t'(\ry,t) \Wy + \Uy (\ry) \) with $\Wy$ constant. Note that $\Uy (\ry)$ cannot be constant, otherwise

\[ \det\left(\frac{\p\ry'}{\p \ry}\right) = \det (\Wy (\na t')^T) = 0, \] 

\n
since every $3\times 3$ matrix of the form $\ay\by^T$ is singular, which would contradict Axiom 5.

Thus \rf{lasec} becomes:

\be \Ky (\ry_0, \vy) = \Wy + \frac{1}{\dof}(\vy\cdot\na)\Uy .\ee{later}

\n
By comparing the functional dependence of the various terms of this equation we get that $\dof$ must be independent of $t$; since

\[  \dof = (\na t')(\ry_0 , t)\cdot\vy + \frac{\p t'}{\p t} (\ry_0 , t) \]

\n
this means that neither $\dss\frac{\p t'}{\p t}$ (put $\vy =\0$), nor $\na t'$ may depend on $t$. It follows that  \( t' = g(\ry) + \al t \) for some $\al >0$ and $g: \R^3 \ra \R$ differentiable. 

{\sl Claim 2}: Both functions $\Uy$ and $g$ are affine.

\n
Equation \rf{later} can be re-written as 

\[ \Ky (\ry_0 , \vy) = \Wy + (\na g(\ry_0)\cdot\vy + \al)^{-1}((\vy\cdot\na )\Uy) (\ry_0).\]

\n
By substituting $\ry_0$ with $\ry (t)$  and differentiating with respect to $t$, we obtain zero; if we neglect the irrelevant denominator and evaluate for $t=t_0$ we get:

\[ \0 = \sum_{\al,\bt=1}^3 \left((\na g(\ry_0)\cdot\vy +\al) \frac{\p ^2\Uy}{\p x^{\al}\p x^{\bt}}(\ry_0)
-\frac{\p ^2 g}{\p x^{\al}\p x^{\bt}}(\ry_0)\frac{\p\Uy}{\p x^{\g}} (\ry_0)v^\g\right)v^{\al}v^{\bt}. \]

By re-arranging the terms we have:

\[ \al\sum_{\al,\bt=1}^3 \frac{\p ^2\Uy}{\p x^\al \p x^\bt}v^\al v^\bt 
=-\sum_{\al,\bt,\g =1}^3 \left(\frac{\p g}{\p x^\g}\frac{\p ^2\Uy}{\p x^\al \p x^\bt} 
- \frac{\p ^2 g}{\p x^\al \p x^\bt}\frac{\p \Uy}{\p x^\g}\right)v^\al v^\bt v^\g. \] 

Now, for every fixed $\ry_0$ at the lefthand side there is a quadratic vector polynomial in the $v^\al$, while at the righthand side there is a cubic polynomial -- both homogeneous with respect to $v^\al$: this is possible, for every $\vy$ in an open neighborhood of $\0$ (Axiom 8), if and only if both polynomials vanish for every $\ry_0$, that is, if all their coefficients are identically zero: 

\be \frac{\p ^2 \Uy}{\p x^\al \p x^\bt} =\0,\;  \frac{\p^2 g}{\p x^\al \p x^\bt}\frac{\p \Uy}{\p x^\g} =\0. \ee{equaff}

\n
From the first equality we have that $\Uy$ is an affine map: $\Uy (\ry) = X\ry + \ay$, so we can write

\[ \left\{\ba{rcl} \ry' &=& (g(\ry) + \al t) \Wy + X\ry + \ay, \\ [4pt] t' &=& g(\ry) + \al t + \ell.\ea\right. \]

\n
Since, as we have seen, $\Uy$ is nonconstant, at least one of the vectors $\dss\frac{\p \Uy}{\p x^\g}$ is nonzero; thus the second equation in \rf{equaff} implies 

\[ \frac{\p^2 g}{\p x^\al \p x^\bt} \equiv 0, \]

\n
which means that also $g$ is an affine map, and Claim 2 is proven; this ends also the proof of the theorem.

\end{document}